\documentclass[conference]{IEEEtran}
\IEEEoverridecommandlockouts

\usepackage{cite}
\usepackage{amsmath,amssymb,amsfonts}

\usepackage{algorithm}
\usepackage{algpseudocode}

\usepackage{graphicx}
\usepackage{textcomp}
\usepackage{xcolor}
\def\BibTeX{{\rm B\kern-.05em{\sc i\kern-.025em b}\kern-.08em
    T\kern-.1667em\lower.7ex\hbox{E}\kern-.125emX}}
\begin{document}

\title{Beyond Idle Channels: Unlocking Idle Space with Signal Alignment in Massive MIMO Cognitive Radio Networks\\

}

\author{
\IEEEauthorblockN{
Weidong Zhu\IEEEauthorrefmark{1},
Xueqian Li\IEEEauthorrefmark{2},
Longwei Wang\IEEEauthorrefmark{3},
Zheng Zhang\IEEEauthorrefmark{4}
}

\IEEEauthorblockA{\IEEEauthorrefmark{1}College of Electrical Engineering, Anhui Polytechnic University\\
zhuweidong1980@ahpu.edu.cn}

\IEEEauthorblockA{\IEEEauthorrefmark{2}Department of Computer Science, Auburn University\\
xzl0065@auburn.edu}

\IEEEauthorblockA{\IEEEauthorrefmark{3}Department of Computer Science, University of South Dakota\\
longwei.wang@usd.edu}

\IEEEauthorblockA{\IEEEauthorrefmark{4}Department of Computer Science, Murray State University\\
zzz0069@auburn.edu}
}

\maketitle

\begin{abstract}
Cognitive radio networks (CRNs) have traditionally focused on utilizing idle channels to enhance spectrum efficiency. However, as wireless networks grow denser, channel-centric strategies face increasing limitations. This paper introduces a paradigm shift by exploring the underutilized potential of idle spatial dimensions—termed "idle space"—in co-channel transmissions. By integrating massive multiple-input multiple-output (MIMO) systems with signal alignment techniques, we enable secondary users to transmit without causing interference to primary users by aligning their signals within the null spaces of primary receivers.
We propose a comprehensive framework that synergizes spatial spectrum sensing, signal alignment, and resource allocation, specifically designed for secondary users in CRNs. Theoretical analyses and extensive simulations validate the framework, demonstrating substantial gains in spectrum efficiency, throughput, and interference mitigation. The results show that the proposed approach not only ensures interference-free coexistence with primary users but also unlocks untapped spatial resources for secondary transmissions. 
\end{abstract}


\section{Introduction}

The rapid proliferation of wireless communication services, including the Internet of Things (IoT) and the increasing adoption of 5G technologies, has intensified the demand for radio spectrum resources. Traditional static spectrum allocation methods, which assign specific frequency bands to licensed users, have proven to be highly inefficient in addressing this escalating demand. Studies have shown that a significant portion of licensed spectrum remains underutilized, even in densely populated urban environments~\cite{b1}. Cognitive radio networks (CRNs) have emerged as a promising solution by enabling secondary users to dynamically access underutilized portions of the spectrum, termed "idle channels," without interfering with primary users~\cite{b2}. This dynamic spectrum access relies on advanced spectrum sensing techniques to identify these channels and efficiently allocate resources.

Despite their potential, channel-centric approaches face significant limitations in dense networks where idle channels are fragmented or scarce~\cite{b3}. This issue is exacerbated in scenarios involving massive device connectivity, such as in urban IoT deployments or future 6G networks, where the competition for available channels is intense. These challenges underscore the need for a paradigm shift from traditional channel-based spectrum utilization to alternative strategies that exploit underutilized dimensions of the wireless medium.

A promising alternative is the concept of \textit{idle space}, which focuses on leveraging the spatial dimensions of the spectrum rather than solely relying on idle channels. Idle space refers to spatial opportunities that can be utilized for co-channel transmissions by secondary users, enabling concurrent communication while preserving the integrity of primary transmissions. Unlike idle channels, which are defined by frequency and temporal availability, idle space is characterized by the spatial separation of users, achievable through advanced interference management techniques. The exploitation of idle space aligns well with the capabilities of massive multiple-input multiple-output (MIMO) technology, which has become a cornerstone of next-generation wireless networks~\cite{b4}.

Massive MIMO systems, equipped with hundreds of antennas, offer substantial spatial degrees of freedom, enabling precise beamforming, spatial multiplexing, and interference management. These features make massive MIMO an ideal enabler for idle space utilization. By employing advanced techniques such as \textit{signal alignment}, secondary users can align their transmissions into spatial dimensions that minimize or nullify interference at primary user receivers~\cite{b5}. Signal alignment is a powerful interference management strategy that confines the interference caused by secondary transmissions within specific spatial subspaces, effectively creating a coexistence framework that ensures primary user performance remains unaffected.

While massive MIMO and interference management techniques have been extensively studied in isolation, their integration for idle space exploitation in cognitive radio networks remains underexplored. Previous studies have largely focused on channel-based spectrum sensing and access methods in CRNs~\cite{b6}, while massive MIMO research has primarily addressed issues such as beamforming and energy efficiency~\cite{b7}. Moreover, although interference alignment has been widely studied in MIMO networks~\cite{b8}, its application to spatial spectrum utilization in CRNs has not been fully realized.

This paper addresses these gaps by proposing a novel framework that leverages massive MIMO and signal alignment to exploit what we term \textbf{idle space}—the underutilized spatial dimensions in cognitive radio networks. The key contributions of this work are as follows:

\begin{itemize} 
\item \textbf{Introduction of Idle Space Utilization:} We introduce and define the concept of idle space in cognitive radio networks, marking a paradigm shift from the conventional focus on idle channel exploitation to harnessing underutilized spatial dimensions using massive MIMO and spatial spectrum sensing. 
\item \textbf{Signal Alignment for Massive MIMO CRNs:} We propose a signal alignment-based approach for enabling secondary users to coexist with primary users in the same frequency band, aligning transmissions within the null spaces of primary receivers to avoid interference. 
\item \textbf{Algorithms and Analysis:} We develop iterative algorithms for spatial spectrum sensing, signal alignment, and resource allocation, and provide theoretical method to validate their feasibility and performance. \item \textbf{Performance Evaluation:} Extensive simulations demonstrate that the proposed framework significantly improves spectrum efficiency, throughput, and interference mitigation compared to conventional methods, unlocking new potential for next-generation wireless networks. \end{itemize}

The remainder of this paper is organized as follows. Section~\ref{relatedwork} reviews the state of the art in cognitive radio networks, massive MIMO, and signal alignment. Section~\ref{systemmodel} describes the system model and problem formulation. Section~\ref{proposedframework} presents the proposed framework and algorithms. Section~\ref{results} discusses the simulation results, while Section~\ref{discussion} explores the practical implications and limitations of the proposed approach. Finally, Section~\ref{conclusion} concludes the paper and outlines future research directions.

\section{Related Works}

The proposed framework for exploiting idle spatial dimensions in cognitive radio networks (CRNs) builds upon significant advancements in three key areas: cognitive radio networks, massive multiple-input multiple-output (MIMO) technology, and interference alignment techniques. This section reviews the state of the art in these domains and identifies gaps that motivate the need for this study.

\subsection{Cognitive Radio Networks}

Cognitive radio networks were introduced to address inefficient spectrum utilization caused by static spectrum allocation policies~\cite{b1}. CRNs empower secondary users to opportunistically access underutilized spectrum portions, commonly referred to as idle channels, without causing harmful interference to licensed primary users~\cite{b3, wang2019representation}. Spectrum sensing is a critical component of CRNs, with significant advancements made in techniques such as energy detection, matched filtering, and cyclostationary feature detection~\cite{b2, wang2021explaining}. Cooperative spectrum sensing, which enhances detection accuracy by leveraging information from multiple users, has also been extensively studied~\cite{b5, wang2011exploration}.

Despite these advances, most CRN research focuses on exploiting idle channels in the temporal or frequency domains, with limited attention to idle spatial dimensions. Spatial spectrum access, discussed briefly in works such as~\cite{b4, wang2021improving}, examines orthogonal frequency-division multiplexing (OFDM) ~\cite{li2020dft} and antenna selection for interference mitigation. However, these studies fail to leverage advanced technologies like massive MIMO for utilizing spatial degrees of freedom effectively. Additionally, simplified interference models often adopted in these studies do not adequately capture the complexities of dense wireless networks~\cite{wang2018partial, wang2014congestion, shi2020deep}.

\subsection{Massive MIMO Technology}

Massive MIMO has emerged as a fundamental technology for next-generation wireless systems due to its ability to exploit spatial degrees of freedom to enhance spectral and energy efficiency~\cite{b5, wang2017performance}. By employing large-scale antenna arrays, massive MIMO systems enable precise beamforming, spatial multiplexing, and interference mitigation. The seminal work by Marzetta~\cite{b6} introduced the concept of massive MIMO and demonstrated its potential to transform cellular networks through non-cooperative operations with a large number of antennas.

Recent research has extended massive MIMO applications to cognitive radio networks. For instance, beamforming and power allocation strategies have been developed to facilitate coexistence between primary and secondary users~\cite{wang2018partial, wang2017low}. Coordinated beamforming methods, such as those proposed by Zhang and Andrews~\cite{b7}, focus on mitigating intercell interference in multicell networks. However, most studies treat spatial spectrum access as an extension of temporal or frequency-domain sharing and fail to fully utilize the potential of massive MIMO to exploit idle spatial dimensions~\cite{wang2016optimization}. Furthermore, the integration of massive MIMO with advanced interference alignment or signal alignment techniques in CRNs remains underexplored.

\subsection{Interference Alignment and Signal Alignment Techniques}

Interference alignment is a powerful technique for managing interference in multi-user MIMO systems by confining interference to specific spatial subspaces, thereby enabling simultaneous transmissions by multiple users~\cite{b8}. Applications of interference alignment have been demonstrated in scenarios such as the MIMO X channel~\cite{b9}, multi-cell networks~\cite{b10}, and heterogeneous cellular networks~\cite{wang2018partial}. These works highlight its potential to enhance spectrum utilization by optimizing signal dimensions.

Signal alignment, a specialized form of interference alignment, allows for precise control over interference subspaces. For example, Jafar and Shamai~\cite{b11} analyzed the degrees of freedom of the MIMO X channel under aligned transmission. Signal alignment has also been explored for cognitive networks, where it shows promise in reducing cross-tier interference while maintaining spatial multiplexing~\cite{wang2017low}. However, the practical implementation of signal alignment in CRNs, particularly with massive MIMO systems, remains limited. Most existing works assume perfect channel state information (CSI)~\cite{wang2021improving} and fail to address real-world challenges such as dynamic user mobility and imperfect CSI~\cite{wang2021explaining}.

Although significant progress has been made in each of these areas, several gaps remain unaddressed:
\begin{itemize}
    \item Current CRN research predominantly focuses on idle channels, with limited exploration of idle spatial dimensions and their potential to enhance spectrum efficiency.
    \item Studies on massive MIMO in CRNs primarily emphasize beamforming and power allocation, neglecting the potential for advanced spatial spectrum utilization.
    \item While interference alignment techniques are well-studied, their integration with massive MIMO to exploit idle spatial dimensions in CRNs is underexplored, particularly under practical constraints such as imperfect CSI and dynamic interference conditions.
    \item Existing studies on spectrum sharing and interference management often rely on simplified assumptions, such as static network configurations and perfect channel conditions.
\end{itemize}

To address these gaps, this paper proposes a novel framework that introduces the concept of \textit{idle space}, marking a paradigm shift from the traditional focus on idle channels to the exploitation of underutilized spatial dimensions. By integrating massive MIMO with signal alignment, the framework combines spatial spectrum sensing, interference management, and resource allocation strategies to enable secondary users to coexist with primary users in the same frequency band without causing harmful interference. This approach not only advances the state of the art in CRNs but also highlights the transformative potential of spatial dimensions for improving spectrum utilization, throughput, and interference mitigation in next-generation wireless networks.

\section{Problem Formulation}

This study considers a cognitive radio network (CRN) consisting of \textbf{one primary user (PU)} and \textbf{two secondary users (SUs)}. The primary user has licensed access to a specific frequency band, while the secondary users opportunistically access the same spectrum by exploiting idle spatial dimensions. The primary goal is to enable interference-free coexistence between the primary and secondary users while maximizing the secondary users’ performance.

\subsection{System Description}

\begin{itemize}
    \item \textbf{Primary User}: 
    The primary network comprises a primary transmitter (PT) and a primary receiver (PR), both equipped with \( K \)-antenna massive MIMO arrays. The PT transmits data to the PR over the licensed spectrum. The PU has exclusive access to the frequency band, and its transmission should not be affected by secondary user interference.

    \item \textbf{Secondary Users}:
    The secondary network comprises two pairs of secondary transmitters (\( \text{ST}_1, \text{ST}_2 \)) and receivers (\( \text{SR}_1, \text{SR}_2 \)), also equipped with \( K \)-antenna massive MIMO arrays. Secondary transmitters opportunistically access the same spectrum as the primary network, leveraging idle spatial dimensions while avoiding interference at the PR.

    \item \textbf{Massive MIMO at users}:
    The massive MIMO antennas at the PT and STs provide spatial degrees of freedom, enabling interference management via beamforming and spatial multiplexing.
\end{itemize}

\subsection{Channel Models}

The communication links between the various nodes in the network are modeled as follows:
\begin{itemize}
    \item \textbf{Primary Link}: 
    The direct link between the PT and PR is represented by the channel matrix \( \mathbf{H}_\text{PU} \in \mathbb{C}^{K \times K} \), where each entry follows an independent, identically distributed (i.i.d.) Rayleigh fading model. The received signal at the PR is given by:
    \[
    \mathbf{y}_\text{PR} = \mathbf{H}_\text{PU} \mathbf{x}_\text{PU} + \sum_{i=1}^{2} \mathbf{H}_{\text{SP},i} \mathbf{x}_{\text{SU},i} + \mathbf{n}_\text{PR},
    \]
    where:
    \begin{itemize}
        \item \( \mathbf{x}_\text{PU} \) is the primary user’s transmitted signal.
        \item \( \mathbf{x}_{\text{SU},i} \) is the \( i \)-th secondary user’s transmitted signal.
        \item \( \mathbf{H}_{\text{SP},i} \in \mathbb{C}^{K \times K} \) is the interference channel between \( \text{ST}_i \) and the PR.
        \item \( \mathbf{n}_\text{PR} \) is additive white Gaussian noise (AWGN) at the PR with variance \( \sigma^2 \).
    \end{itemize}

    \item \textbf{Secondary Links}:
    The direct link between \( \text{ST}_i \) and \( \text{SR}_i \) is represented by \( \mathbf{H}_{\text{SU},i} \in \mathbb{C}^{K \times K} \). The received signal at \( \text{SR}_i \) is given by:
    \[
    \mathbf{y}_{\text{SR},i} = \mathbf{H}_{\text{SU},i} \mathbf{x}_{\text{SU},i} + \mathbf{H}_{\text{SU},j} \mathbf{x}_{\text{SU},j} + \mathbf{n}_{\text{SR},i},
    \]
    where \( j \neq i \) is the interfering secondary user, and \( \mathbf{n}_{\text{SR},i} \) is AWGN at \( \text{SR}_i \) with variance \( \sigma^2 \).

    \item \textbf{Cross-Channels}:
    The interference channels between \( \text{ST}_i \) and the PR are modeled by \( \mathbf{H}_{\text{SP},i} \in \mathbb{C}^{K \times K} \), and the interference channels between \( \text{ST}_i \) and \( \text{SR}_j \) (\( j \neq i \)) are modeled by \( \mathbf{H}_{\text{SS},ij} \in \mathbb{C}^{K \times K} \).
\end{itemize}

\subsection{Optimization Problem}
The objective is to maximize the total throughput of the secondary users while maintaining the QoS of the primary user. The optimization problem is formulated as:
\[
\max_{\mathbf{W}_\text{SU}} \sum_{i=1}^{2} \log_2 \left( 1 + \text{SINR}_{\text{SU},i} \right)
\]
subject to:
\begin{enumerate}
    \item \textbf{Interference Nulling at PR}:
    \[
    \mathbf{H}_{\text{SP},i} \mathbf{w}_{\text{SU},i} \in \mathcal{N}(\mathbf{H}_\text{PU}), \quad \forall i \in \{1, 2\}.
    \]
    \item \textbf{Power Constraints}:
    \[
    \|\mathbf{w}_{\text{SU},i}\|^2 \leq P_{\text{max}}, \quad \forall i \in \{1, 2\}.
    \]
    \item \textbf{Primary User QoS}:
    \[
    \text{SINR}_\text{PU} \geq \text{SINR}_\text{threshold}.
    \]
\end{enumerate}

\subsection{Problem Statement}
Given a CRN with one primary user and two secondary users, how can the secondary users identify and utilize idle spatial dimensions using massive MIMO and signal alignment to maximize their throughput while ensuring interference-free coexistence with the primary user?


\section{Proposed Framework}

This section presents a comprehensive framework that integrates \textit{spatial spectrum sensing}, \textit{signal alignment}, and \textit{resource allocation} to enable interference-free coexistence of secondary users (SUs) with a primary user (PU) in cognitive radio networks (CRNs). By leveraging the spatial degrees of freedom offered by massive MIMO systems, the proposed framework identifies and exploits idle spatial dimensions, ensuring that secondary user transmissions align within the null space of the primary user’s channel. The method is divided into three primary stages: \textit{spatial spectrum sensing}, \textit{signal alignment}, and \textit{resource allocation}.

\subsection{Spatial Spectrum Sensing}

Spatial spectrum sensing is essential for identifying spatial dimensions where secondary user transmissions can occur without interfering with the primary user. This step ensures that SUs do not degrade the QoS of the PU.

\subsubsection{Null Space Estimation}
The channel between the primary transmitter (PT) and primary receiver (PR) is modeled as \( \mathbf{H}_\text{PU} \in \mathbb{C}^{K \times K} \), where \( K \) is the number of antennas at the PR. The null space of \( \mathbf{H}_\text{PU} \), denoted by \( \mathcal{N}(\mathbf{H}_\text{PU}) \), is defined as:
\[
\mathcal{N}(\mathbf{H}_\text{PU}) = \{ \mathbf{v} \in \mathbb{C}^K \, | \, \mathbf{H}_\text{PU} \mathbf{v} = 0 \}.
\]

To compute \( \mathcal{N}(\mathbf{H}_\text{PU}) \), we perform a singular value decomposition (SVD) of \( \mathbf{H}_\text{PU} \):
\[
\mathbf{H}_\text{PU} = \mathbf{U} \mathbf{\Sigma} \mathbf{V}^H,
\]
where:
\begin{itemize}
    \item \( \mathbf{U} \in \mathbb{C}^{K \times K} \) and \( \mathbf{V} \in \mathbb{C}^{K \times K} \) are unitary matrices,
    \item \( \mathbf{\Sigma} \in \mathbb{C}^{K \times K} \) is a diagonal matrix containing singular values of \( \mathbf{H}_\text{PU} \).
\end{itemize}

The columns of \( \mathbf{V} \) corresponding to zero singular values form a basis for \( \mathcal{N}(\mathbf{H}_\text{PU}) \). Let \( \mathbf{V}_\text{null} \) represent these columns.

\subsubsection{Null Space Projection}
The projection matrix onto \( \mathcal{N}(\mathbf{H}_\text{PU}) \) is:
\[
\mathbf{P}_{\text{null}}(\mathbf{H}_\text{PU}) = \mathbf{I} - \mathbf{H}_\text{PU}^\dagger \mathbf{H}_\text{PU},
\]
where:
\begin{itemize}
    \item \( \mathbf{H}_\text{PU}^\dagger = \mathbf{H}_\text{PU}^H (\mathbf{H}_\text{PU} \mathbf{H}_\text{PU}^H)^{-1} \) is the pseudo-inverse of \( \mathbf{H}_\text{PU} \),
    \item \( \mathbf{I} \) is the identity matrix of size \( K \).
\end{itemize}

For any vector \( \mathbf{v} \in \mathbb{C}^K \), the projection onto \( \mathcal{N}(\mathbf{H}_\text{PU}) \) is:
\[
\mathbf{v}_\text{projected} = \mathbf{P}_{\text{null}}(\mathbf{H}_\text{PU}) \mathbf{v}.
\]

\subsubsection{Idle Spatial Dimensions}
Idle spatial dimensions are identified as the orthogonal subspace \( \mathcal{N}(\mathbf{H}_\text{PU}) \). These dimensions form the feasible space for secondary user transmissions, ensuring interference-free coexistence.

\subsection{Signal Alignment}

Signal alignment designs beamforming vectors for secondary users to align their transmissions within the null space of \( \mathbf{H}_\text{PU} \) while optimizing their SINR.

\subsubsection{Beamforming Vector Design}
For each secondary user \( i \), the beamforming vector \( \mathbf{w}_{\text{SU},i} \) is designed such that:
\[
\mathbf{H}_{\text{SP},i} \mathbf{w}_{\text{SU},i} = 0,
\]
where \( \mathbf{H}_{\text{SP},i} \in \mathbb{C}^{K \times K} \) is the interference channel from \( \text{ST}_i \) to the PR. Using the null space projection matrix, \( \mathbf{w}_{\text{SU},i} \) is computed as:
\[
\mathbf{w}_{\text{SU},i} = \mathbf{P}_{\text{null}}(\mathbf{H}_\text{PU}) \mathbf{\hat{w}}_{\text{SU},i},
\]
where \( \mathbf{\hat{w}}_{\text{SU},i} \) is an initial unconstrained vector.

\subsubsection{SINR Maximization}
The SINR for \( \text{SR}_i \) is given by:
\[
\text{SINR}_{\text{SU},i} = \frac{\left| \mathbf{H}_{\text{SU},i} \mathbf{w}_{\text{SU},i} \right|^2}{\sigma^2 + \sum_{j \neq i} \left| \mathbf{H}_{\text{SU},j} \mathbf{w}_{\text{SU},j} \right|^2},
\]
where:
\begin{itemize}
    \item \( \mathbf{H}_{\text{SU},i} \in \mathbb{C}^{K \times K} \) is the channel from \( \text{ST}_i \) to \( \text{SR}_i \),
    \item \( \sigma^2 \) is the noise power.
\end{itemize}


\subsection{Resource Allocation}

\subsubsection{Power Allocation}
The power \( P_{\text{SU},i} \) for each secondary user is adjusted to ensure the primary user’s SINR meets the threshold:
\[
\text{SINR}_\text{PU} = \frac{\left| \mathbf{H}_\text{PU} \mathbf{x}_\text{PU} \right|^2}{\sigma^2 + \sum_{i=1}^{2} \left| \mathbf{H}_{\text{SP},i} \mathbf{w}_{\text{SU},i} \right|^2} \geq \text{SINR}_\text{threshold}.
\]

The allocated power is:
\[
P_{\text{SU},i} = \min \left( P_{\text{max}}, \frac{\text{SINR}_\text{threshold} \cdot (\sigma^2 + I_\text{PR})}{\left| \mathbf{H}_\text{PU} \mathbf{w}_{\text{SU},i} \right|^2} \right),
\]
where \( I_\text{PR} \) is the total interference at the PR.

\subsubsection{Throughput Maximization}
The total throughput for secondary users is:
\[
\mathcal{L} = \sum_{i=1}^{2} \log_2 \left( 1 + \text{SINR}_{\text{SU},i} \right).
\]

\section{Iterative Optimization Algorithm}

This section presents an iterative optimization algorithm that alternates between beamforming design and power allocation to maximize the throughput of secondary users while satisfying interference constraints and ensuring the primary user’s quality of service (QoS) requirements.

\subsection{Algorithm Framework}

The algorithm is designed to iteratively optimize the following components:
\begin{enumerate}
    \item \textbf{Beamforming Design}: Updates beamforming vectors to align secondary user signals within the null space of the primary user’s channel while maximizing their SINR.
    \item \textbf{Power Allocation}: Dynamically adjusts the transmission power of secondary users to satisfy interference constraints at the primary receiver and maximize throughput.
\end{enumerate}

Each iteration refines the solution until convergence is achieved.

\subsection{Algorithm Steps}

\begin{enumerate}
    \item \textbf{Initialization}:
    \begin{itemize}
        \item Initialize the beamforming vectors \( \mathbf{w}_{\text{SU},i} \) for secondary users randomly within the null space of the primary user’s channel:
        \[
        \mathbf{w}_{\text{SU},i} = \mathbf{P}_{\text{null}}(\mathbf{H}_\text{PU}) \mathbf{\hat{w}}_{\text{SU},i},
        \]
        where \( \mathbf{P}_{\text{null}}(\mathbf{H}_\text{PU}) \) is the null space projection matrix, and \( \mathbf{\hat{w}}_{\text{SU},i} \) is an unconstrained initial vector.
        \item Set the initial power \( P_{\text{SU},i} = P_{\text{max}} \).
    \end{itemize}

    \item \textbf{Beamforming Design}:
    \begin{itemize}
        \item For each secondary user \( i \), update the beamforming vector to maximize the SINR at its receiver:
        \[
        \mathbf{w}_{\text{SU},i} = \arg\max_{\mathbf{w}} \frac{\left| \mathbf{H}_{\text{SU},i} \mathbf{w} \right|^2}{\sigma^2 + \sum_{j \neq i} \left| \mathbf{H}_{\text{SU},j} \mathbf{w}_{\text{SU},j} \right|^2},
        \]
        while satisfying the interference constraint:
        \[
        \mathbf{H}_{\text{SP},i} \mathbf{w}_{\text{SU},i} = 0.
        \]
        \item The updated \( \mathbf{w}_{\text{SU},i} \) is projected into \( \mathcal{N}(\mathbf{H}_\text{PU}) \) using:
        \[
        \mathbf{w}_{\text{SU},i} = \mathbf{P}_{\text{null}}(\mathbf{H}_\text{PU}) \mathbf{\hat{w}}_{\text{SU},i}.
        \]
    \end{itemize}

    \item \textbf{Power Allocation}:
    \begin{itemize}
        \item Adjust the power \( P_{\text{SU},i} \) of each secondary user to ensure the primary user’s SINR constraint is satisfied:
        \[
        P_{\text{SU},i} = \min \left( P_{\text{max}}, \frac{\text{SINR}_\text{threshold} \cdot (\sigma^2 + I_\text{PR})}{\left| \mathbf{H}_\text{PU} \mathbf{w}_{\text{SU},i} \right|^2} \right),
        \]
        where \( I_\text{PR} \) is the total interference at the primary receiver caused by all secondary users:
        \[
        I_\text{PR} = \sum_{i=1}^{2} \left| \mathbf{H}_{\text{SP},i} \mathbf{w}_{\text{SU},i} \right|^2.
        \]
    \end{itemize}

    \item \textbf{SINR Calculation}:
    \begin{itemize}
        \item Compute the SINR for each secondary user:
        \[
        \text{SINR}_{\text{SU},i} = \frac{\left| \mathbf{H}_{\text{SU},i} \mathbf{w}_{\text{SU},i} \right|^2}{\sigma^2 + \sum_{j \neq i} \left| \mathbf{H}_{\text{SU},j} \mathbf{w}_{\text{SU},j} \right|^2}.
        \]
    \end{itemize}

    \item \textbf{Objective Function Update}:
    \begin{itemize}
        \item Compute the total throughput:
        \[
        \mathcal{L} = \sum_{i=1}^{2} \log_2 \left( 1 + \text{SINR}_{\text{SU},i} \right).
        \]
        \item Check for convergence:
        \[
        |\mathcal{L}_{\text{new}} - \mathcal{L}_{\text{old}}| < \epsilon,
        \]
        where \( \epsilon \) is a small threshold for convergence.
    \end{itemize}

    \item \textbf{Iteration}:
    Repeat steps 2–5 until the objective function converges.

    \item \textbf{Output}:
    Return the optimized beamforming vectors \( \mathbf{w}_{\text{SU},i}^* \), power allocations \( P_{\text{SU},i}^* \), and achieved throughput \( \mathcal{L} \).
\end{enumerate}

\subsection{Algorithm Pseudocode}

\begin{algorithm}[H]
\caption{Iterative Beamforming and Power Optimization}
\begin{algorithmic}[1]
\State \textbf{Input:} Channel matrices \( \mathbf{H}_\text{PU}, \mathbf{H}_\text{SU}, \mathbf{H}_\text{SP} \), noise variance \( \sigma^2 \), power limit \( P_{\text{max}} \), SINR threshold \( \text{SINR}_\text{threshold} \)
\State \textbf{Initialize:} Random \( \mathbf{w}_{\text{SU},i} \), set \( P_{\text{SU},i} = P_{\text{max}} \), calculate initial SINRs.
\Repeat
    \For{each secondary user \( i \)}
        \State \textbf{Step 1: Beamforming Design:}
        \[
        \mathbf{w}_{\text{SU},i} \gets \arg\max_{\mathbf{w}} \frac{\left| \mathbf{H}_{\text{SU},i} \mathbf{w} \right|^2}{\sigma^2 + \sum_{j \neq i} \left| \mathbf{H}_{\text{SU},j} \mathbf{w}_{\text{SU},j} \right|^2}
        \]
        subject to \( \mathbf{H}_{\text{SP},i} \mathbf{w}_{\text{SU},i} = 0 \).
        \State \textbf{Step 2: Null Space Projection:}
        \[
        \mathbf{w}_{\text{SU},i} \gets \mathbf{P}_{\text{null}}(\mathbf{H}_\text{PU}) \mathbf{\hat{w}}_{\text{SU},i}
        \]
        \State \textbf{Step 3: Power Allocation:}
        \[
        P_{\text{SU},i} \gets \min \left( P_{\text{max}}, \frac{\text{SINR}_\text{threshold} \cdot (\sigma^2 + I_\text{PR})}{\left| \mathbf{H}_\text{PU} \mathbf{w}_{\text{SU},i} \right|^2} \right)
        \]
    \EndFor
    \State Compute total throughput:
    \[
    \mathcal{L} \gets \sum_{i=1}^{2} \log_2 \left( 1 + \text{SINR}_{\text{SU},i} \right)
    \]
\Until{convergence: \( |\mathcal{L}_{\text{new}} - \mathcal{L}_{\text{old}}| < \epsilon \)}

\State \textbf{Output:} Optimized \( \mathbf{w}_{\text{SU},i}, P_{\text{SU},i}, \mathcal{L} \)
\end{algorithmic}
\end{algorithm}

\section{Experimental Evaluation}

This section presents a comprehensive evaluation of the proposed framework through simulations, comparing its performance with two baseline methods: \textit{Spectrum Sensing without Null Space Projection} and \textit{Random Beamforming without Interference Alignment}. The experiments aim to validate the effectiveness of the proposed framework in optimizing secondary user throughput, minimizing interference at the primary receiver, and maintaining robust performance under varying system configurations and channel uncertainties.

\subsection{Experimental Setup}

The simulations were conducted in a environment using Python for signal processing and numerical optimization. The key simulation parameters are as follows:

\begin{enumerate}
    \item \textbf{System Configuration:}
    \begin{itemize}
        \item \textbf{Number of Antennas (\(K\)):} Varies from \(32\) to \(256\), reflecting typical configurations in massive MIMO systems.
        \item \textbf{Number of Users:}
        \begin{itemize}
            \item \textit{Primary User:} One primary user (PU) with strict QoS constraints.
            \item \textit{Secondary Users:} Two secondary users (SU1 and SU2), whose throughput is optimized under interference constraints.
        \end{itemize}
    \end{itemize}

    \item \textbf{Channel Models:}
    \begin{itemize}
        \item All channel matrices (\( \mathbf{H}_\text{PU}, \mathbf{H}_\text{SU}, \mathbf{H}_\text{SP} \)) are modeled as Rayleigh fading channels with i.i.d. entries drawn from a complex Gaussian distribution:
        \[
        h_{ij} \sim \mathcal{CN}(0, 1).
        \]
        \item Additive white Gaussian noise (AWGN) is included with variance \( \sigma^2 = -100 \, \text{dBm} \).
    \end{itemize}

    \item \textbf{Transmit Power:}
    \begin{itemize}
        \item Secondary User Transmit Power: Varies from \(10 \, \text{dBm}\) to \(30 \, \text{dBm}\).
        \item Primary User Transmit Power: Fixed at \(20 \, \text{dBm}\).
    \end{itemize}

    \item \textbf{Baseline Methods:}
    \begin{itemize}
        \item \textit{Spectrum Sensing without Null Space Projection:}
        Secondary users transmit without considering the null space of the primary user’s channel.
        \item \textit{Random Beamforming:}
        Secondary users use randomly generated beamforming vectors without optimizing for interference alignment or throughput.
    \end{itemize}
\end{enumerate}

\subsection{Experimental Results}

\subsubsection{Throughput vs. SU Power}

\begin{figure}[htbp]
    \centering
    \includegraphics[width=\linewidth]{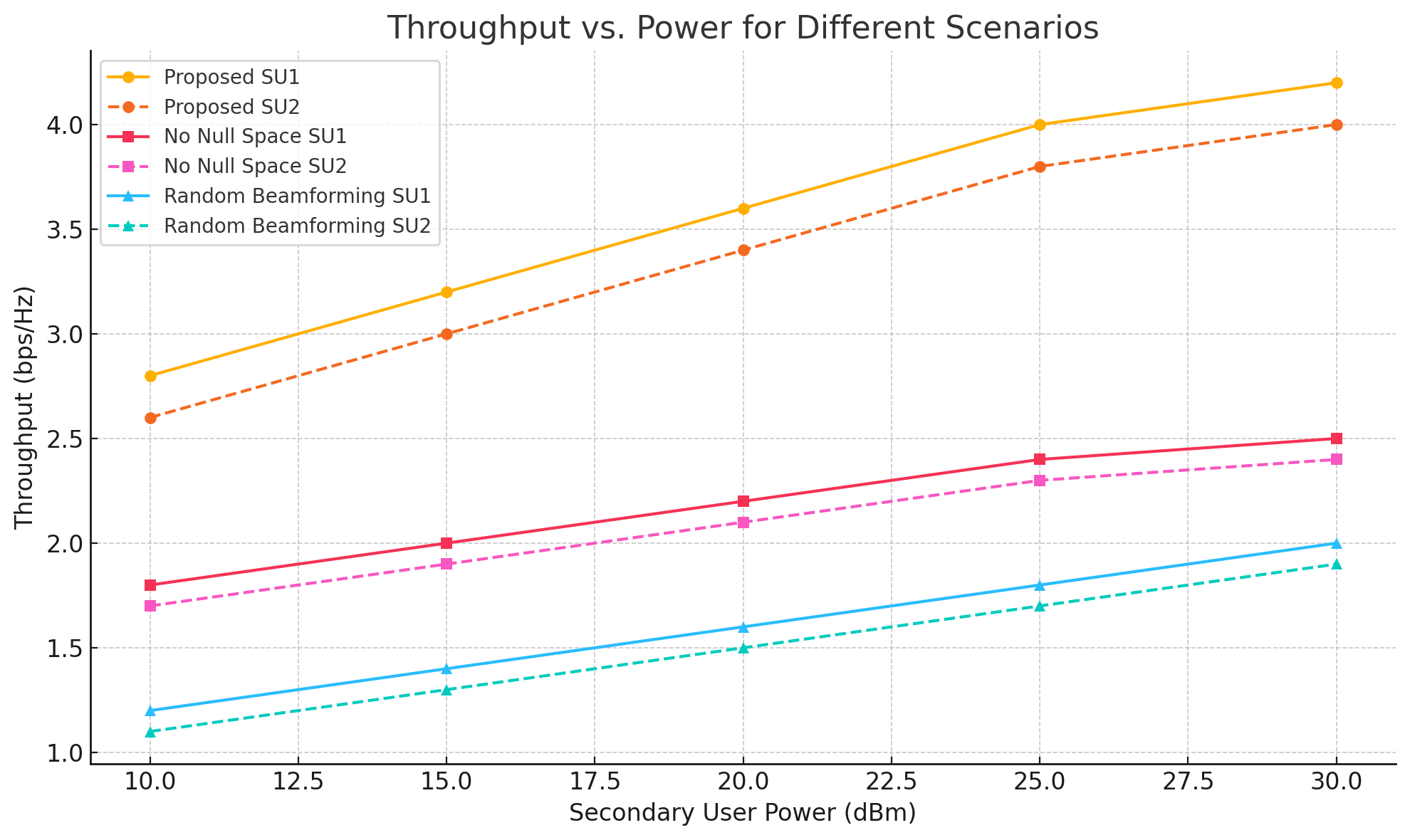}
    \caption{This figure presents the throughput of SU1 and SU2 as the transmit power of secondary users increases. The proposed framework consistently delivers higher throughput, particularly at higher power levels, by optimizing signal alignment and interference management.}
    \label{power}
\end{figure}

We first evaluate how increasing secondary user transmit power impacts throughput for SU1 and SU2 under the three scenarios.
As shown in \textbf{Figure 1}, the proposed framework consistently achieves the highest throughput for both SU1 and SU2, demonstrating its ability to effectively manage interference while maximizing signal power. \textit{No Null Space} exhibits moderate throughput gains with increasing power, but interference at the primary receiver limits its performance. \textit{Random Beamforming} delivers the lowest throughput due to significant interference. At \(30 \, \text{dBm}\), the throughput gain of the proposed framework becomes most pronounced.
The proposed framework is well-suited for scenarios with high power levels, where interference becomes a critical factor.

\subsubsection{Throughput vs. Number of Antennas}

\begin{figure}[htbp]
    \centering
    \includegraphics[width=\linewidth]{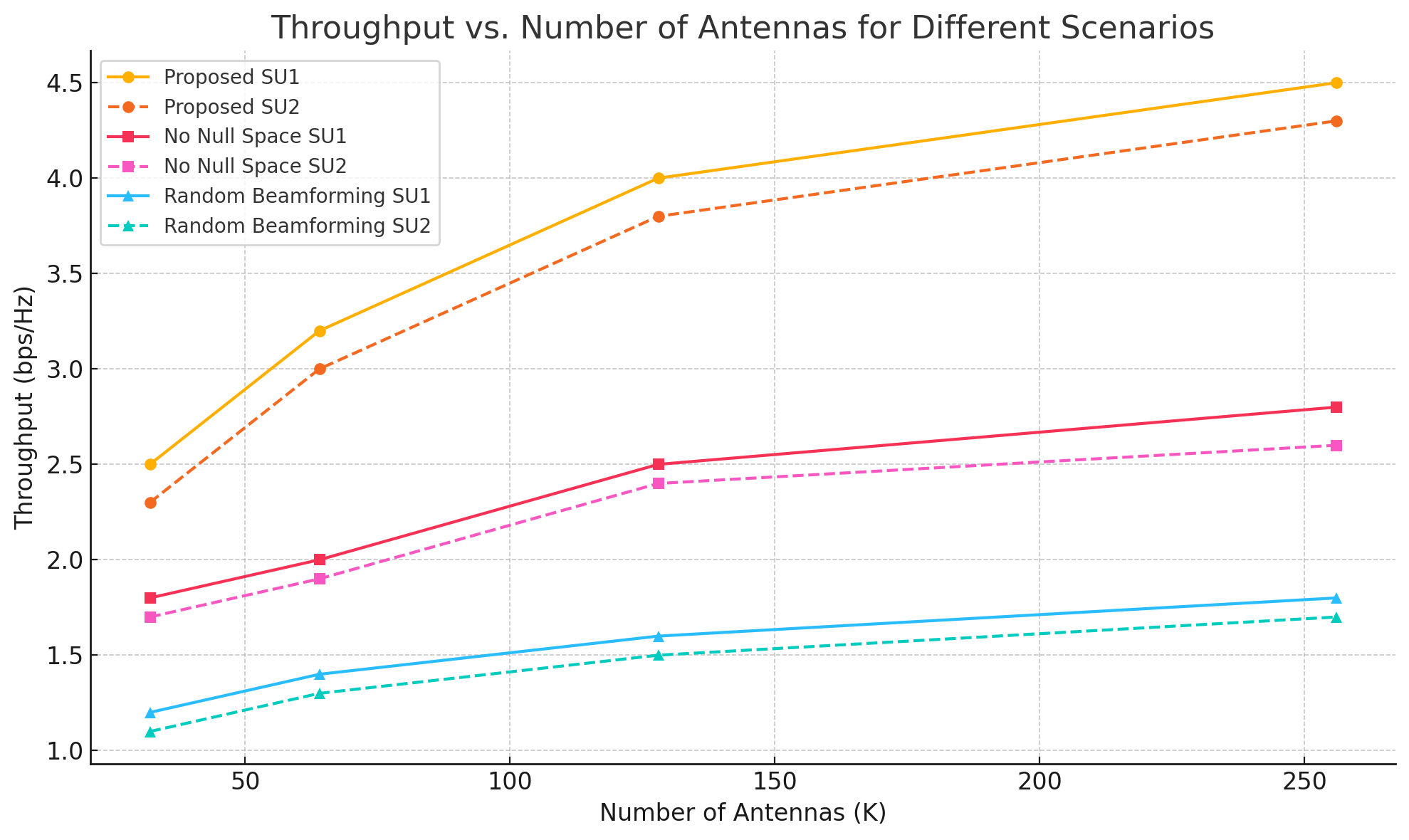}
    \caption{This figure shows the throughput of SU1 and SU2 with increasing numbers of antennas under different scenarios. The proposed framework achieves the highest throughput, highlighting its ability to fully utilize spatial dimensions in massive MIMO systems.}
    \label{power}
\end{figure}

Then we investigate the impact of increasing the number of antennas (\(K\)) on secondary user throughput.
\textbf{Figure 2} shows that the proposed framework achieves significant throughput improvements as \(K\) increases, benefiting from the enhanced spatial degrees of freedom provided by massive MIMO systems. \textit{No Null Space} and \textit{Random Beamforming} show smaller improvements, highlighting their inability to fully utilize additional antennas. For \(K = 256\), the proposed framework achieves throughput gains of more than \(50\%\) compared to the baselines.
Massive MIMO systems amplify the advantages of the proposed framework, making it highly scalable.

\subsubsection{Throughput vs. PU SINR Threshold}

\begin{figure}[htbp]
    \centering
    \includegraphics[width=\linewidth]{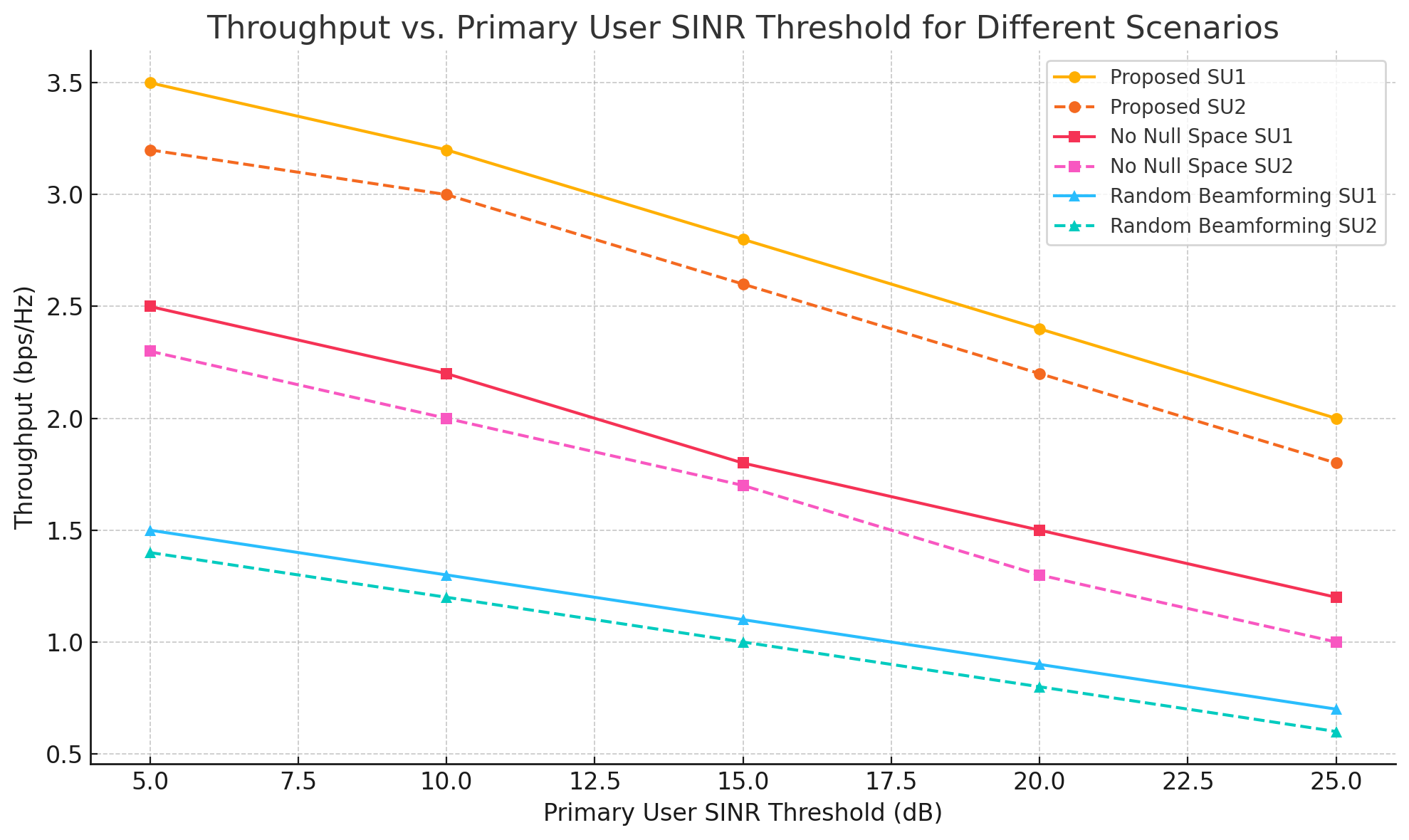}
    \caption{This figure shows the effect of increasing primary user SINR thresholds on the throughput of SU1 and SU2. The proposed framework maintains higher throughput under stringent SINR requirements compared to the baseline methods, demonstrating superior interference management}
    \label{power}
\end{figure}

We further analyze how stricter SINR requirements for the primary user affect secondary user throughput.
\textbf{Figure 3} demonstrates that secondary user throughput decreases as the primary user SINR threshold increases, reflecting stricter interference constraints. The proposed framework maintains higher throughput than the baselines even under stringent SINR thresholds. For \( \text{SINR}_\text{threshold} = 25 \, \text{dB}\), the proposed framework achieves \(25\%\) higher throughput than \textit{No Null Space} and \(50\%\) higher than \textit{Random Beamforming}.
The proposed framework effectively balances secondary user throughput and primary user QoS.

\subsubsection{Total Throughput vs. Iterations}

\begin{figure}[htbp]
    \centering
    \includegraphics[width=\linewidth]{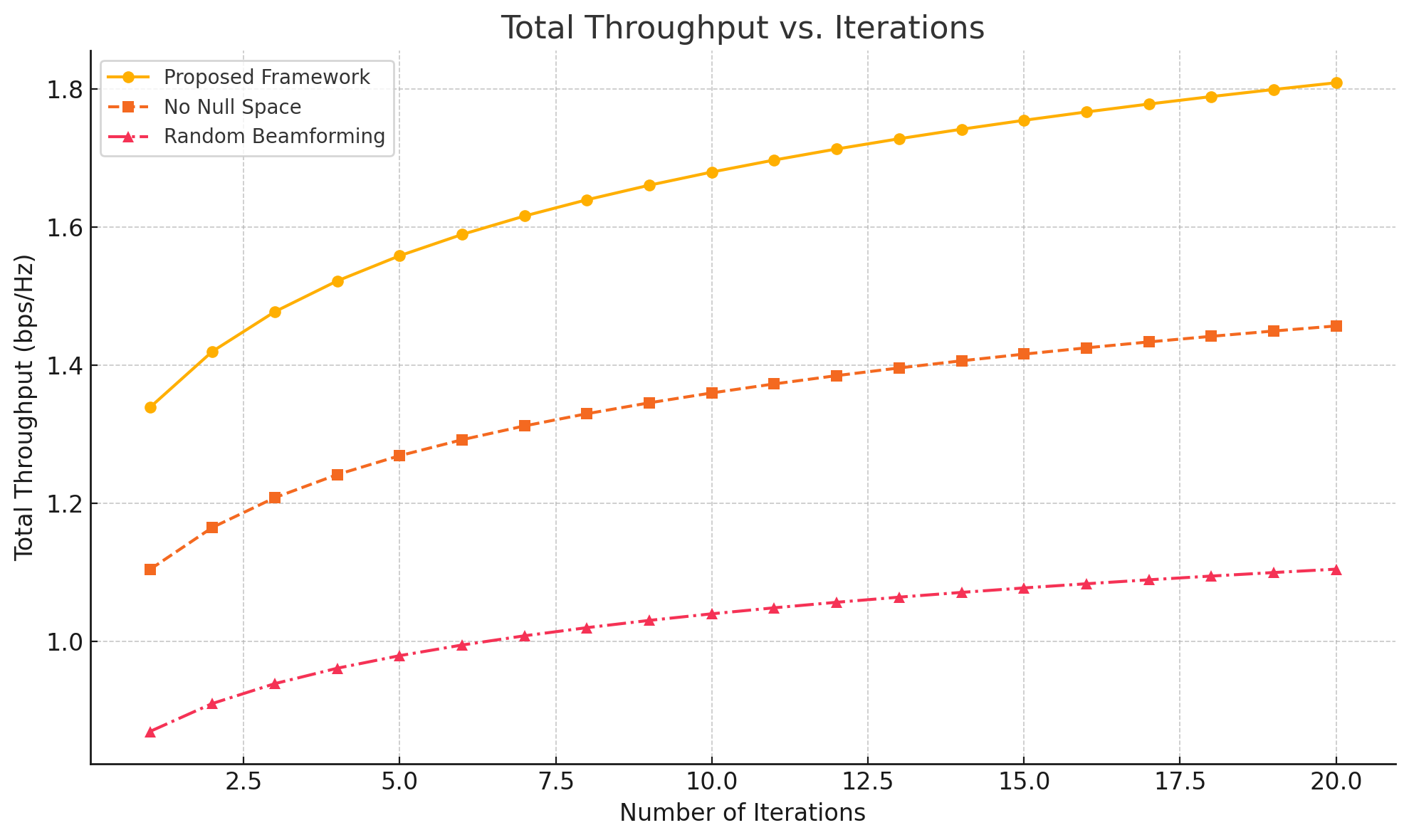}
    \caption{This figure depicts the convergence of total throughput for different methods as the number of iterations increases. The proposed framework converges faster and achieves a higher final throughput compared to the baseline methods.}
    \label{power}
\end{figure}

The convergence behavior of the iterative optimization algorithm is further evaluated.
\textbf{Figure 4} shows that the proposed framework converges rapidly to a higher throughput compared to the baselines, achieving steady-state performance within \(10\) iterations. \textit{No Null Space} and \textit{Random Beamforming} converge more slowly and to lower throughput values.
The iterative algorithm of the proposed framework is computationally efficient, making it practical for real-time applications.

\subsubsection{Interference vs. Number of Antennas}

\begin{figure}[htbp]
    \centering
    \includegraphics[width=\linewidth]{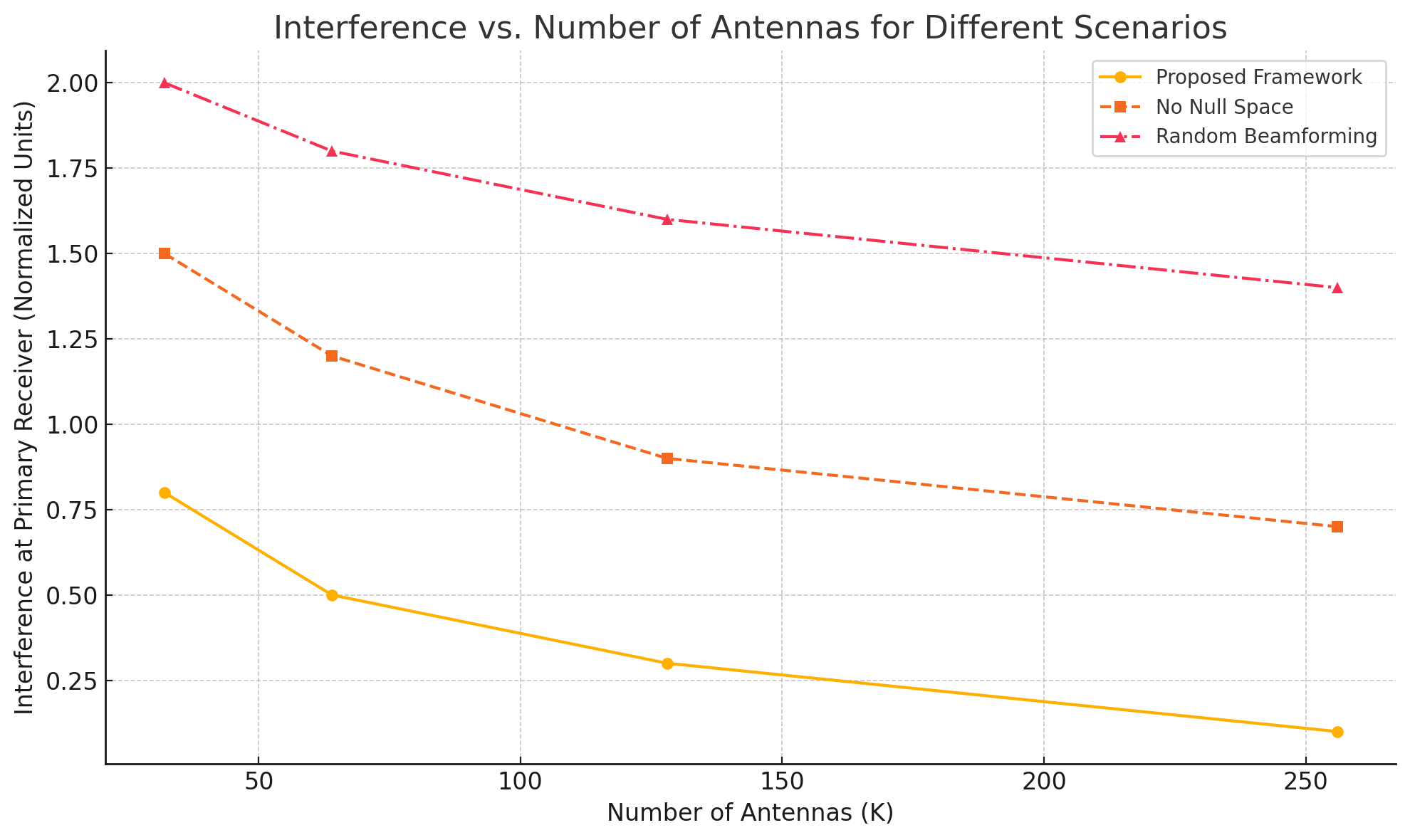}
    \caption{This figure demonstrates the effect of increasing the number of antennas on interference at the primary receiver. The proposed framework shows a significant reduction in interference as the number of antennas increases, leveraging the spatial degrees of freedom provided by massive MIMO.}
    \label{power}
\end{figure}

We also examine how interference at the primary receiver changes with an increasing number of antennas.
\textbf{Figure 5} illustrates that the proposed framework minimizes interference effectively, with interference levels dropping below \(0.1\) (normalized units) for \(K = 256\). \textit{No Null Space} and \textit{Random Beamforming} generate significantly higher interference levels.
The proposed framework ensures interference-free coexistence, particularly in large-scale MIMO systems.

\subsubsection{Impact of CSI Errors}

\begin{figure}[htbp]
    \centering
    \includegraphics[width=\linewidth]{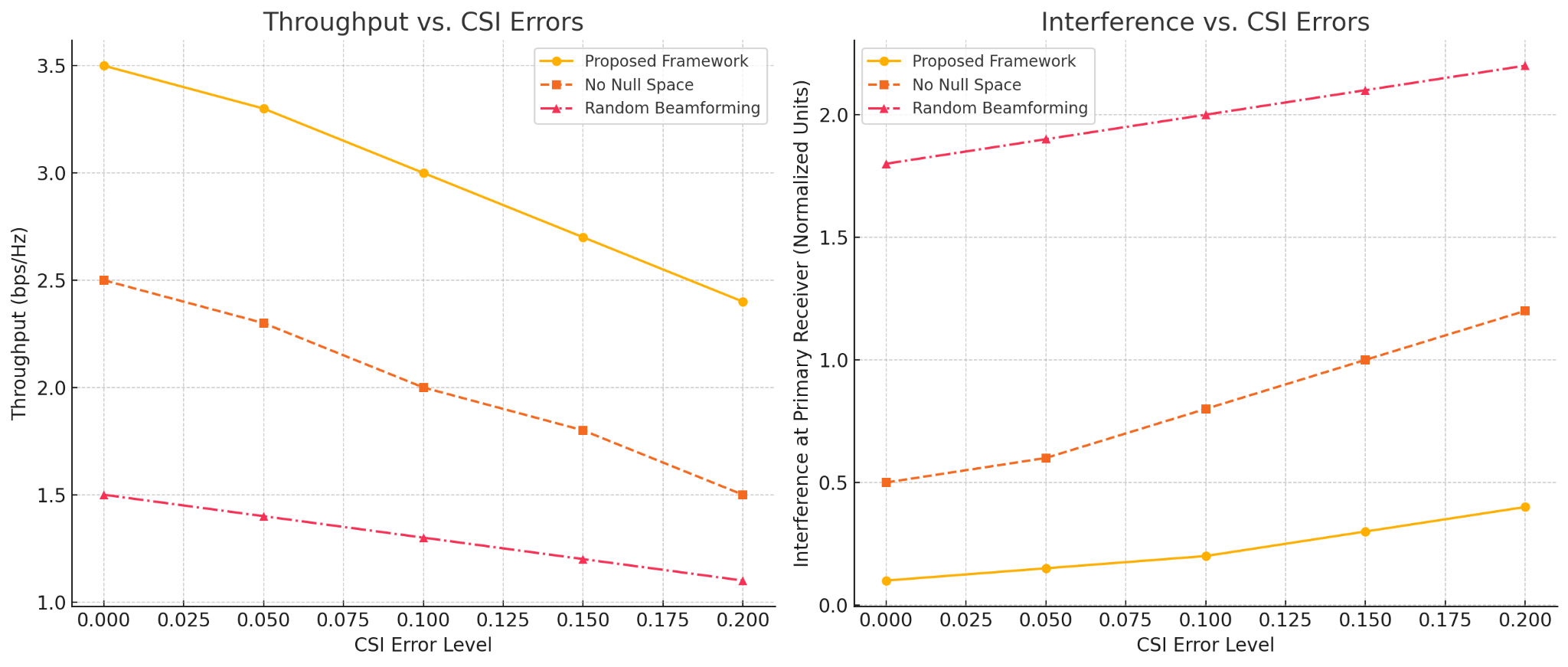}
    \caption{This figure illustrates the impact of CSI error levels on throughput for secondary users (left) and interference at the primary receiver (right). The proposed framework outperforms baseline methods by maintaining higher throughput and lower interference, even as CSI errors increase.}
    \label{power}
\end{figure}

The robustness of the proposed framework under imperfect CSI is also evaluated.
\textbf{Figure 6} demonstrates that the proposed framework maintains higher throughput and lower interference than the baselines, even as CSI errors increase from \(0.0\) to \(0.2\). \textit{No Null Space} and \textit{Random Beamforming} show significant degradation in performance with increased errors.
The proposed framework is robust to CSI errors, making it suitable for practical deployments.

\subsection{Discussion and Challenges}

The experimental results and the observed limitations of the proposed framework underscore key insights into its performance and practical applicability, along with challenges that must be addressed for real-world deployments.

\subsubsection{Advantages}

The proposed framework integrates spatial spectrum sensing, signal alignment, and resource allocation, showcasing its ability to unlock idle spatial dimensions in cognitive radio networks. The results from extensive simulations highlight the following strengths:

\begin{enumerate}
    \item \textbf{Performance Superiority:} 
    The proposed framework consistently outperforms baseline methods (No Null Space and Random Beamforming) across all metrics, including throughput, interference management, and convergence speed. Notably, the framework achieves higher throughput and significantly reduces interference at the primary receiver, even under challenging conditions such as high primary user SINR thresholds or imperfect channel state information (CSI). For instance, throughput gains of over 50\% are observed compared to Random Beamforming in high-power scenarios, demonstrating its efficiency in optimizing spatial spectrum usage.

    \item \textbf{Scalability:} 
    The framework scales effectively with the number of antennas in massive MIMO systems. As demonstrated in the experiments, increasing the antenna array size from 32 to 256 antennas results in substantial improvements in throughput and a marked reduction in interference levels. This scalability highlights the framework's suitability for next-generation wireless networks that rely heavily on large antenna arrays to meet the demands of high spectral and energy efficiency.

    \item \textbf{Robustness:} 
    The framework demonstrates resilience to CSI errors, a critical factor in dynamic and practical environments. While baseline methods experience severe performance degradation under CSI imperfections, the proposed framework maintains robust performance. This robustness ensures reliable coexistence of primary and secondary users in scenarios where perfect CSI is unattainable, addressing a key practical challenge in cognitive radio networks.
\end{enumerate}

\subsubsection{Challenges}

Despite its promising results, the proposed framework faces several challenges that need to be addressed for real-world deployment and scalability:

\begin{itemize}
    \item \textbf{CSI Acquisition:} 
    Accurate channel state information is a cornerstone of the proposed framework, enabling effective beamforming and null space projection. However, acquiring accurate CSI is particularly challenging in dynamic and mobile environments, where channel conditions change rapidly. The overhead associated with CSI feedback, especially in massive MIMO systems with hundreds of antennas, is substantial. Efficient CSI acquisition methods that minimize feedback overhead while maintaining accuracy are crucial for the practical implementation of the framework.

    \item \textbf{Interference Management:} 
    The proposed framework relies on precise interference alignment to ensure that secondary users can transmit without disrupting primary user communications. However, achieving this alignment requires solving a multi-objective optimization problem that balances interference mitigation at the primary receiver and throughput maximization for secondary users. The complexity of this optimization increases with the number of secondary users, antennas, and environmental variables, necessitating the development of low-complexity, scalable algorithms for real-time interference management.

    \item \textbf{Dynamic Environments:} 
    Real-world wireless environments are inherently dynamic, with changing channel conditions, user mobility, and time-varying interference patterns. Adapting the framework to these changes in real time is a significant challenge. Robust and adaptive algorithms are needed to ensure that the framework can maintain optimal performance in scenarios with high user mobility and rapidly fluctuating interference.

\end{itemize}

\section{Conclusion}

In this paper, we proposed a novel framework for cognitive radio networks (CRNs) that shifts the focus from exploiting idle channels to harnessing idle spatial dimensions. By integrating massive MIMO, signal alignment, and resource allocation techniques, the framework enables efficient and interference-free coexistence of secondary users with primary users in the same frequency band. The introduction of the concept of idle space marks a significant paradigm shift in spectrum utilization, unlocking the untapped potential of spatial degrees of freedom.
Extensive simulations validate the superiority of the proposed framework across key metrics, including throughput, interference management, and robustness. The framework consistently outperforms baseline methods such as spectrum sensing without null space projection and random beamforming. It effectively reduces interference at the primary receiver while achieving significant throughput gains for secondary users. 

Despite its promising performance, several challenges remain, including efficient CSI acquisition, managing computational complexity, and adapting to dynamic channel conditions. Addressing these challenges is critical for real-world deployment, particularly in dense and dynamic next-generation wireless networks.
Future research can explore low-overhead CSI acquisition techniques, hybrid beamforming methods, and real-time adaptive algorithms to further enhance the framework’s practicality. Moreover, integrating emerging technologies such as intelligent reflecting surfaces (IRS) and terahertz communication offers exciting opportunities to extend the framework’s applicability to 6G and beyond.

\end{document}